\documentclass[aps,preprint]{revtex4}

\usepackage[dvips]{graphicx}

\newcommand{\heff}{H_{\rm eff}}

\begin{document}

\date{\today}
\title{
Dynamical stabilization and time in open quantum systems}

\author{Ingrid Rotter\footnote{rotter@pks.mpg.de}}
\address{Max-Planck-Institut f\"ur Physik komplexer
Systeme, D-01187 Dresden, Germany 
%\\ rotter@pks.mpg.de
}

\begin{abstract}
The meaning of time in an open quantum system is considered under 
the assumption that both, system and environment, are quantum 
mechanical objects. The Hamilton operator of the system is 
non-Hermitian. Its imaginary part is the time operator.
As a rule, time and energy vary continuously when 
controlled by a parameter.  At high 
level density, where many states avoid crossing,
a dynamical phase transition takes place in the 
system under the influence of the environment. It causes a
dynamical stabilization of the system
what can be seen in many different experimental data.
Due to this effect,
time is bounded from below: the decay widths (inverse proportional to 
the lifetimes of the states) do not increase limitless. 
The dynamical stabilization is an irreversible process.

\end{abstract}

\maketitle

\section{Introduction}
\label{1}

The problem of time is considered in very many papers starting from 
the very beginning of quantum mechanics. For example,
the derivation of the uncertainty relation between space and momentum
can be found in every textbook on quantum mechanics. However, the
uncertainty relation between energy and time could  not be derived 
convincingly up to now.   According to Pauli \cite{pauli}, 
the reason is that energy is bounded from below while this is not the
case for time and, furthermore, time varies continuously in contrast
to energy. A critical consideration of the discussions of the time 
energy uncertainty relation  is given in 
\cite{aharonov}. The authors point out that energy $E$ characterizes 
the quantum system while time $t$ is measured by an external clock.

This point of view is followed up in recent studies \cite{briggs1}: 
{\it it is important in the application of a given 
time energy uncertainty relation to state precisely what kind of 
measurement is being made 
and to specify accordingly the meaning of the $\Delta t$ involved,
e.g. does it refer to the accuracy of measurement, to the duration 
of measurement or perhaps to the lifetime of a decaying state}. 
Also the derivation of the time dependent Schr\"odinger equation is 
considered  \cite{briggs2}: {\it This derivation, unlike
those presented in quantum mechanics textbooks but in the spirit of
Schr\"odinger's original approach to the problem, acknowledges that
time enters quantum mechanics only when an external force on the 
quantum system is considered classically.} Starting from a fully 
time-independent formulation of quantum mechanics, it is possible
systematically to derive the time-dependent Schr\"odinger 
equation for a quantum system in the approximation that the 
environment is treated semiclassically. The derivation relies on 
the assumption of a closed object comprising system plus environment
\cite{briggs2}.

In all these considerations the Hamiltonian of the system is assumed
to be Hermitian, and the lifetime of a decaying state is calculated
perturbatively. The situation is another one when both, system and
environment, are quantum mechanical objects, i.e. the environment
is the continuum of scattering wavefunctions into which the
system is embedded. In such a case, the
Hamilton operator of the system is non-Hermitian and the lifetime of
a decaying state is calculated non-perturbatively \cite{top}.   

It is the aim of the present paper to consider the meaning of time
in an open quantum system described by a non-Hermitian operator 
the eigenvalues of which are complex. The basic equations are
taken from the review \cite{top}. It turns out that a natural
definition of time in an open quantum system is related to the decay
widths of the states and, furthermore, to the dynamical stabilization
occurring in the system under the influence of the environment. Time,
defined in such a manner, is bounded from below. 

The paper is organized in the following manner. In section
\ref{2} it is shown  why in an open quantum system with many levels 
the non-Hermitian operator  $\heff$ appears and how it looks like.  
Using this model, the properties of discrete and narrow resonance 
states are sketched in section \ref{3}. In both cases, 
exceptional points play an important role. 
Since there is much confusion in the literature on 
these singular points,  their properties are sketched in the appendixes
\ref{eiv} to \ref{a5} from a unified point of view and compared to 
the results of experimental studies. The relation of the
imaginary part of the eigenvalues of $\heff$ to time
is considered in section \ref{4} while 
in the following section \ref{5} the decay rate is calculated.
The phenomenon of a dynamical phase transition in an open quantum 
system is explained in section \ref{6}.
In section \ref{7}, a few examples of dynamical phase transitions,
that are found  experimentally, are given. In any case, they cause a
dynamical stabilization of the system.
Based on these results, it is possible to define  {\it
  time} in an open quantum system and to discuss its properties 
(section \ref{8}). The results are summarized in the last section \ref{9}.

\section{Non-Hermitian Hamilton operator of an open quantum system}
\label{2}

The definition of an {\it  open quantum  system} used in the present paper 
is the following: a quantum system is
considered to be open when it is embedded into the environment of the
continuum of scattering wavefunctions. According to this definition, 
the system is  localized while the environment is extended
infinitely. This type of embedding of a 
quantum system into an environment {\it always} exists. The 
environment may be changed by external forces, e.g. by a laser in the 
case of atoms (examples are shown in \cite{marost}).  
It can however {\it not} be deleted completely.
When allowed by the  energy of the system 
(and no special selection rules hold), most states of the system
decay into the continuum of scattering wavefunctions and have a finite 
lifetime. Otherwise they are discrete.

A method being very suitable for the description of this situation, is a
projection operator formalism with projection onto system and environment, 
respectively.  
In the following, some basic equations of such a 
formalism are given, for details see the review \cite{top}.
First the Schr\"odinger equations in the two subspaces
(with the Hamiltonian $H_B$ and $H_c$, respectively) have to be solved
and the projection operators $Q$ and $P$ have to be defined,
\begin{eqnarray}
(H_{B} - E_k^{B}) \, \Phi_k^B = 0 \quad & \to & \quad
Q  =  \sum_{k =1}^N|\Phi_k^B\rangle\langle \Phi_k^B | 
\label{form1} \\
(H_{c} - E) \, \xi^{E}_{c} = 0 \quad & \to & \quad
P  =  \sum_{c=1}^C\int_{-\infty}^{\infty} dE \; 
|\xi^{E}_{c} \rangle \langle \xi^{E}_{c} | \; .
\label{form2}
\end{eqnarray}
The $\Phi_k^B$ are the wavefunctions describing the $N$ discrete states of
the closed many-particle system, while the $\xi_c^E$ are the scattering
wavefunctions of the environment consisting of $C$ continua
($\xi^{E}_{c}$  is written instead of $\xi^{E(+)}_{c}$ for convenience).
The Schr\"odinger equation in the whole function space with
discrete and scattering states reads
\begin{eqnarray}
(H^{\rm full} -E)~\Psi_c^E  =  0
\label{formfull}
\end{eqnarray}
with the Hermitian operator 
$H^{\rm full}  =  H_{QQ} + H_{QP} + H_{PQ} + H_{PP}$ and $Q+P=1$
where $H_{QQ} \equiv QHQ$ and so on.  
The coupling matrix elements between system and environment are
\begin{eqnarray}
\gamma_{k c}^0  = 
\sqrt{2\pi}\, \langle \Phi_k^B| H_{QP} | \xi^{E}_{c}
\rangle \; .
\label{form3}
\end{eqnarray}
The solution of the full problem (\ref{formfull}) is 
\begin{eqnarray}
\Psi^E_c =   \xi^E_c  ~+ \sum_{k=1}^N
\Omega_k \cdot  \frac{\langle \Phi_k^*| H_{QP} | 
\xi^E_{c} \rangle }
{E - z_k}
\label{form4}
\end{eqnarray}
with the non-Hermitian operator
\begin{eqnarray}
{H}_{\rm eff}\; =\; H_{QQ} + H_{QP} G_P^{(+)} H_{PQ} 
\;\equiv \; H_{B} + V_{BC} G_C^{(+)} V_{CB} 
\label{form5}
\end{eqnarray}
and
\begin{eqnarray}
(H_{\rm eff}-z_k)~\Phi_k=0 \; ; \qquad z_k\equiv E_k - \frac{i}{2}\; \Gamma_k 
\label{form6}
\end{eqnarray}
after diagonalization. Here
$E_k=E_k(E)$ and $\Gamma_k=\Gamma_k(E)$ are, respectively,  
the position in energy and the decay width (inverse lifetime) 
of the state $k$  at the energy $E$, see equations
(\ref{formfull}) and (\ref{form4}).

Position  and width  of the resonance
state $k$ are energy independent numbers $E_k=E_k^p$ and $\Gamma_k=\Gamma_k^p$ 
only when the state $k$ is not overlapped by another resonance state 
and, furthermore, it is far from a decay
threshold. These numbers  can be obtained also,
as usually, from the poles of the $S$ matrix, see \cite{varvara}.
The functions $z_k=z_k(E)$ describe, in any case, the line shape of 
resonances correctly. 
Further, $G_P^{(+)} = P (E - H_{PP})^{-1} P$ is the 
Green function in the $P$ subspace and
$\Omega_k =(1+ G_{P}^{(+)} \; H_{PQ}) \, \Phi_k$
is the wavefunction of the resonance state. 
The eigenfunctions of $\heff$ are biorthogonal,
$\langle \Phi_k^* | \Phi_l \rangle  =\delta_{k,l} $
and $\langle \Phi_k | \Phi_k \rangle  \equiv A_k \ge 1 $. 

The Hamiltonian $\heff$ consists {\it formally}
of a first-order and a second-order interaction term.
The second-order term via the continuum 
determines the dynamics of the system at high level density. It
leads to the principal value integral
\begin{eqnarray}
{\rm Re}\; 
\langle \Phi_i^{B} | {H}_{\rm eff} |  \Phi_j^{B} \rangle 
 -  E_i^B \delta_{ij} 
=\frac{1}{2\pi} \sum_{c=1}^C {\cal P} 
\int_{\epsilon_c}^{\epsilon_{c}'} 
 {\rm d} E' \;  
\frac{\gamma_{ic}^0 \gamma_{jc}^0}{E-E'} 
\label{form11}
\end{eqnarray}
and the residuum 
\begin{eqnarray}
{\rm Im}\; \langle \Phi_i^{B} | {H}_{\rm eff} |
  \Phi_j^{B} \rangle =
- \frac{1}{2}\; \sum_{c=1}^C  \gamma_{ic}^0 \gamma_{jc}^0 \; .
\label{form12}
\end{eqnarray}
When $i=j$, the expression 
Re$\langle \Phi_i^{B} | {H}_{\rm eff} |  \Phi_j^{B} \rangle$
gives the shift in energy of the state $i$ due to the interaction between the
state and the large system it is part of, i.e. due to the interaction of the
state $i$ with the environment (subspace $P$). This is the {\it self-energy}
of the state which is analog to the Lamb shift known in atomic
physics.   When $i\ne j$,  (\ref{form11}) describes the 
energy shift of the state $i$ due to its coupling to another state
$j \ne i$ via the environment (continuum of scattering wavefunctions). 
These couplings to the different states $j$ cause, at high level
density, collectively contributions to the 
energy shift of the state $i$. In atomic physics, 
contributions of such a type  are studied
newly experimentally and called collective Lamb shift 
\cite{roehlsberger,scully}.

The energy window coupled directly to the continuum is 
$\epsilon_c \le E \le \epsilon_{c}'$. In nuclei $\epsilon_{c}' \to \infty$
while $\epsilon_c$ denotes the lowest threshold for emission of a particle.
The energy shift  $\Delta_k = E_k - E_k^B $, including the corresponding
corrections arising from the coupling of different states via the continuum,
can {\it not} be simulated by
two-body forces \cite{mix}. It is a global property, see \cite{jung}, and 
is {\it not} contained in any standard calculation with a Hermitian 
Hamilton operator for the many-body problem. 

The method sketched here for the description of an open quantum system
is not the only one. 
The advantage of the explicit consideration of $H_{\rm eff}$ 
in the model described above,
consist, above all, in the fact that {\it the many-body problem 
in the $Q$ subspace has to be solved only once}, since it is 
energy independent, see (\ref{form1}). At low level density, 
${H}_{\rm eff} \approx H_{B} \equiv H_{QQ}$, the second-order term can be 
treated perturbatively, i.e. the Hamiltonian $H_{\rm eff}$ is almost Hermitian,
corresponding to the assumption of standard  quantum physics.
At high level density, however, the second-order term can {\it not} be treated
perturbatively. It induces a global mixing of the states according to 
(\ref{form11}) \cite{jung}, 
and deviations from standard  quantum physics will occur.
They appear most clearly when the number of decay channels (continua)
is small, especially for $C=1$.

\section{Discrete and narrow resonance states}
\label{3}

The eigenvalues of the effective Hamiltonian $\heff$, equation
(\ref{form5}), may be real or complex. In the first case, the
eigenstates are discrete states  while they are resonance states in the
second case. The
boundary conditions are different for the two different types of states.
In both cases, the trajectories of the eigenvalues 
traced as a function of any control parameter avoid crossing, usually. The
corresponding crossing points (called mostly {\it exceptional points})
of two eigenvalue trajectories can be found by analytical continuation.
They play an important role for the dynamics of open
quantum systems. The  eigenvalues and eigenfunctions at and in the
neighborhood of an exceptional point (including the experimental proof
of their main features) are given in 
appendix \ref{eiv} and \ref{eif}, respectively.

{\it Resonance states} are 
coupled directly to the continuum,
$\epsilon_c \le E_k \le \epsilon_{c}'$.  The eigenvalues $z_k$ of $\heff$
are complex, generally. 
According to (\ref{form4}), the  scattering wavefunction 
$\Psi^{E}_{c~{\rm int}}$ inside the system can be represented
in the set $\{\Phi_k\}$ of the biorthogonal eigenfunctions of $\heff$,
i.e. 
\begin{eqnarray}
\Psi^E_{c ~{\rm int}}=\sum_{k =1}^N 
\frac{\langle  \Phi_{k}^*| H_{QP} | \xi^E_{c} \rangle }{E - z_k}
~\Phi_k \, .
\label{add1}
\end{eqnarray}
The expression $\langle \Phi_k^* | \Phi_l \rangle$ is
a complex number, and a consistent normalization of the  $\Phi_k$ requires 
${\rm Im}\langle \Phi_k^* | \Phi_k \rangle$ $= 0$. This corresponds to
some rotation such that the  phases of the eigenfunctions relative to 
one another are not rigid (when traced as a function of a certain parameter).
Instead, it holds $0\le \rho \le 1$ (for details see \cite{top}) for
the phase rigidity 
\begin{eqnarray}
\rho = 
e^{2i\theta} \frac {\int dr ([{\rm Re} \Psi^{E}_{c~{\rm int}}]^{2} -
[{\rm Im} \Psi^{E}_{c~{\rm int}}]^2)} 
{\int dr ([{\rm Re} \Psi^{E}_{c~{\rm int}}]^2 +
[{\rm Im} \Psi^{E}_{c~{\rm int}}]^2)} 
\label{add2}
\end{eqnarray}
where $\theta$ is the rotation angle (see appendix \ref{a3} for the
definition of the phase rigidity $r_k$ in the case of a two-level system). 
Only at low level density
and far from avoided level crossings, 
$\langle \Phi_k^* | \Phi_l \rangle \approx \langle \Phi_k | \Phi_l \rangle $
and $\rho \approx 1$ such that
standard Hermitian quantum physics is a good approximation.
Otherwise, spectroscopic redistribution processes take place,
$\rho <1$, the Sch\"odinger equation has a nonlinear source term (see
appendix \ref{a4}) and time reversal symmetry is broken (see appendix
\ref{a5}). Here, the description of the system
by standard Hermitian quantum mechanics breaks down.

The breakdown of standard quantum mechanics can be understood in the
following manner.  When $1 > \rho > 0$ the states with the wavefunctions 
$ \Psi^{E}_{c~{\rm int}}$ avoid crossing (when controlled by a parameter)
and become mixed globally
such that a few states of the system can 
align {\it hierarchically} (i.e. step by step \cite{top}) 
with the scattering states $\xi^E_c$ of  the 
environment. As a result,  their decay widths $\Gamma_k$ become large.
Full alignment is reached for $\rho = 0$. The
alignment of a few states $\Psi^{E}_{c~{\rm int}}$ with the channel
wavefunction $\xi^E_c $ occurs by trapping  other
resonance states, i.e. by (partial or complete) decoupling them  
from the environment.  This is nothing but width bifurcation:
the widths of a few states become large while the widths of the other
ones become small by varying the control parameter. This scenario occurs
in the vicinity of  exceptional points \cite{top,jopt}.
When $\rho < 1$, the system can not be described perturbatively, 
and standard Hermitian quantum physics fails (compare Appendixes \ref{a3}
to \ref{a5} for the two-level system).
Similar results are obtained by using other methods, e.g. 
\cite{past,leve,schom,reichl}.

The influence of the continuum of scattering wavefunctions ($P$
subspace) onto the {\it discrete states} of the system with
energies $E_k$  beyond the energy window 
$\epsilon_c \le E \le \epsilon_{c}'$ , seems to
be much less important. For discrete states $\heff$ is non-Hermitian, 
but the eigenvalues $z_k=E_k$ are real, the eigenfunctions $\Phi_k$ are
orthogonal, $\langle \Phi_i | \Phi_{k}\rangle = 
\delta_{i k} $, and the phase rigidity is $\rho =1$. However, 
$E_k \ne E_k^B$. According to (\ref{form11}), the energy shift $\Delta_k 
= E_k - E_k^B \ne 0$ is caused by the coupling of the state $k$ to the 
environment,  i.e. by the embedding of the system into the continuum
of scattering  wavefunctions. By this,  many-body forces  are induced
in the system  \cite{mix}. Discrete states avoid crossing and, 
at a critical value of the control parameter, the two states are exchanged
as known for about 80 years \cite{landau}. The only difference to the
avoided crossing of resonance states is that discrete states never
cross. The corresponding crossing point can be found only by
analytical continuation into the continuum \cite{solov}.
 
When the level density is high, many discrete
levels avoid crossing (when traced as a function of a  
control parameter). For illustration let us consider an $A$ particle system.
In this case, the induced many-body forces 
cause a global mixing of the discrete states in a finite parameter range,
see \cite{jung}. 
Finally, an {\it aligned} discrete state is formed with the structure
{\it bound particle +  (A-1) particle 
residual system} which corresponds to the structure of the decay channel
{\it unbound particle +  (A-1) particle residual system}
(the quantum numbers of  particle and  residual system are
the same in both cases 
and (A-1) denotes the number of particles of the localized residual 
system after emission of one  particle into the continuum). 
This discrete state is the analog to an aligned resonance state
above particle decay threshold. The only difference between these two
states is that the energy of the preformed aligned discrete state
is too small and  does not allow the emission of one particle from the
system,  while the  
decay width of the aligned resonance state is large, corresponding to a 
short lifetime of this state. 

It should be mentioned that energy conservation is not the only source
for an eigenvalue of $\heff$ to be real. Im($z_k) \equiv -\Gamma_k
/2=0 $ is possible also due to selection rules (according to the 
corresponding quantum numbers of the states), or because of width 
bifurcation appearing  at high  level density 
(causing the so-called bound states in the continuum \cite{top}).
Another source for real eigenvalues of a non-Hermitian operator  
appears in PT symmetric systems (where P and T denote parity and time, 
respectively) \cite{bender} and can be observed 
in optics due to the formal equivalence of
the optical wave equation in PT symmetric optical lattices to the
quantum mechanical Schr\"odinger equation \cite{opticsexp}.
The relation of these results 
to the properties of open quantum systems as discussed
above, is considered in \cite{jopt}, see also appendix \ref{eiv}.

\section{The imaginary part of the eigenvalues
of the operator $\heff$}
\label{4}

In order to study the physical meaning of Im$(z_k) = - \Gamma_k/2$, equation
(\ref{form6}), the behavior of three neighboring resonances in a
two-dimensional quantum billiard connected to a single waveguide 
is investigated theoretically in \cite{persson}. 
A measurable quantity derived from the reflection coefficient $R(E)$
is the {\it Wigner Smith time delay function} 
\begin{eqnarray}
\tau_w=\frac{d\Theta}{dE} \; .
\label{im2}
\end{eqnarray}
It is the time the wave spends inside the billiard.
The energies $E_k^p$ and widths $\Gamma_k^p$ of the resonance states
can be found from the {\it poles of the function $R(E)$} analytically
continued into the lower complex plane. 

In \cite{persson}, contour and surface plot of $ln(\tau_w)$ and the
motion of the corresponding three resonance poles by varying the coupling
between the waveguide and the resonator are  calculated.
At weak coupling to the waveguide, the
three resonances are seen clearly. As the coupling to the waveguide
increases, the lifetimes of all three states decrease, as expected. 
As the resonances start to influence one another, the states
attract each other in energy, two
of them become trapped while the third one becomes short-lived. At
further increasing coupling, the lifetimes of the two trapped
resonance states increase, contrary to expectation. 
The lifetime of the short-lived state is,
at large opening, so short that it practically disappears when
plotting the time delay. The motion of the poles is reflected
in the time delay function. The calculations have shown further that the 
interference between the resonance states leads to a
mixing of their wavefunctions with respect to the eigenfunctions of
the closed resonator (defined by decoupling the resonator from the waveguide). 

Some years ago, the dynamics of resonance states is studied
experimentally by means of
a flat microwave resonator connected to a waveguide where the coupling
strength between resonator and wave\-guide can be varied by hand
\cite{stm}. In this experiment, the microwaves 
enter the billiard through a slit, the opening of which can be varied. The 
motion of the resonance poles as a function of the opening of the slit  
is traced starting close to the real axis and following them into the
region of overlapping resonances. The  results verify  
the resonance trapping effect discussed above. This experimental proof
does not depend on any model assumptions. Meanwhile, the resonance
trapping effect has been investigated and verified in many other
theoretical and experimental studies performed on different systems 
(see section \ref{7} and review \cite{top}).

As a result of these studies, the physical meaning of the
imaginary part Im$(z_k)$  of the eigenvalues
of the non-Hermitian Hamilton operator $\heff$ is directly
related to the time $\tau_w$. This relation holds also at high level density
where the system as a whole is dynamically stabilized (see section \ref{6}).

\section{Decay rate at high level density}
\label{5}
 
The time dependent Schr\"odinger equation for the scattering
wavefunctions $\Psi^E_{c ~{\rm int}}$ inside the system reads
\begin{eqnarray}
H^{\rm full} ~\Psi^E_{c ~{\rm int}}(t)=
i\hbar \frac{\partial}{\partial t} ~\Psi^E_{c ~{\rm int}}(t) 
\label{tdse0}
\end{eqnarray}
with $H^{\rm full}$ defined in (\ref{formfull}) and $\Psi^E_{c ~{\rm int}}$
in (\ref{add1}). The right solutions $|\Psi^E_{c ~{\rm int}}\rangle$
may be represented by an ensemble of resonance states 
$k$ that describes the decay
of the localized part of the system  at the energy $E$,
 \begin{eqnarray} 
|\Psi^E_{c ~{\rm int}}(t)\rangle   =
e^{-iH_{\rm eff} \, t/\hbar} ~|\Psi^E_{c ~{\rm int}}(t_0)\rangle 
= \sum_{k=1}^N ~e^{-i z_{k} \, t/\hbar}  
c_{c\, k}^E ~|\Phi_{k }\rangle 
\label{tdse1}
\end{eqnarray}
with   $c_{c\, k}^E = \langle\Phi_k^* |V|\xi^E_c\rangle/(E-z_k) $
according to (\ref{add1}). 
By means of (\ref{tdse1}) and the corresponding expression for the
left  solution  of (\ref{tdse0}), the population probability 
$\langle \tilde\Psi_c(t)|\tilde\Psi_c(t)\rangle = 
\sum_\lambda c_{c\, k}^{~2}  ~e^{- \Gamma_{k } t/\hbar} $
with the energy averaged values $c_{c\, k}$ can be defined. The 
decay rate reads \cite{decayrate}
\begin{eqnarray} 
k_{\rm gr}(t) = - \frac{\partial}{\partial t} ~{\rm ln} 
\, \langle \Psi^E_{c ~{\rm int}}(t) | \Psi^E_{c ~{\rm int}}(t) \rangle 
= \frac{1}{\hbar}~\frac{\sum_k \Gamma_\lambda ~c_{c\, k}^{~2}
~e^{- \Gamma_{k } t/\hbar}}{\sum_k 
c_{c\, k}^{~2}  ~e^{- \Gamma_{k } t/\hbar}} \; .
\label{tdse4}
\end{eqnarray}
The decay properties of the resonance states
can be studied best when their excitation 
takes place in a time interval that is very short 
as compared to the lifetime $\tau_\lambda$ of the resonance states.
In such a case, no perturbation of the decay process
by the still continuing excitation process will take place.

For an isolated  resonance state $k$,  (\ref{tdse4}) 
passes into the standard expression
$k_{\rm gr}(t) ~\to ~k_k ~= ~\Gamma_{k }/\hbar \, . $
In this case, the quantity $k_k$ is constant in time and  corresponds to 
the standard relation $\tau_k = \hbar / \Gamma_k$ with $\tau_k = 1/k_k $.
It describes the idealized case with exponential decay law and
a Breit-Wigner resonance in the cross section.

Equation (\ref{tdse4}) describes, however, the decay rate also in the
regime of overlapping resonances \cite{decayrate}. The overlapping and
mutual influence of resonance states is maximal at the 
avoided (and true) crossing points in
the complex plane  where two  eigenvalues $z_{k}$ and
$z_{k '}$ of the effective Hamilton operator $H_{\rm eff}$  coalesce
(or almost coalesce). Nevertheless, the decay rate is everywhere
smooth as can be seen also directly from (\ref{tdse4}).
This result coincides with the general statement 
according to which all observable
quantities behave smoothly at  singular points. 

An interesting result is the saturation of the average decay rate 
$k_{\rm av}$ in the regime of strongly overlapping resonances. 
According to the bottle-neck picture of the transition state theory, 
it starts at a certain critical value of bound-continuum coupling 
\cite{miller,millerresp}. As has been shown in \cite{comment}, 
this is caused by width bifurcation 
since the definition of an average lifetime of the 
resonance states is meaningful  only for either the 
long-lived states or the short-lived ones. 
The  widths $\Gamma_k$ of the long-lived (trapped) states  
are almost the same for all  the different states $k$, i.e. 
$\Gamma_{\rm av} \approx \Gamma_k$ for all long-lived 
resonance states \cite{top}. It follows therefore 
$k_{\rm av} \approx \Gamma_{\rm av} /\hbar$ 
from (\ref{tdse4}). According to the average width $\Gamma_{\rm av}$,
the average lifetime  of the long-lived states can be defined by 
$\tau_{\rm av} = 1/k_{\rm av} $. Then
$\tau_{\rm av} = \hbar / \Gamma_{\rm av}$. That means, the basic relation
between lifetimes and decay widths of resonance states holds 
not only for isolated resonance states 
but also for the narrow (trapped) resonance states, i.e. for 
$\Gamma_{\rm av}$ and $\tau_{\rm av}$.

\section{Resonance trapping and dynamical phase transitions}
\label{6}

Some years ago, the question has been studied \cite{jung}  whether or not
the resonance trapping phenomenon is related to some type of phase 
transition. The study is performed by using the toy model 
\begin{eqnarray}
H_{\rm eff}^{\rm toy} = H_0 + i \alpha VV^+
\label{sol15}
\end{eqnarray}
where $H^0$ and $VV^+$ are Hermitian, $V$ is the coupling vector of the
system to the environment and the parameter $\alpha$ simulates 
the coupling strength between system and environment.
The calculations are carried out for
the one-channel case and with the assumption that (almost) all 
crossing (exceptional) points  accumulate in one point \cite{hemuro}.
The control parameter $\alpha$ is a real number.
It has been found that  resonance trapping  
may be understood, in this case, as a second-order phase transition.
The calculations are performed for a linear chain consisting of a
finite number $N$ of states. The state in the center of the spectrum 
traps the other ones and becomes a collective state in a global
sense: it contains components of almost all basic states of the
system, also of those which are not overlapped by it. The normalized width 
$\Gamma_0/N$ of this state can be considered as the order parameter:  
it increases linearly as a function of $\alpha$, and the first
derivative of $\Gamma_0/N$ jumps at the critical value 
$\alpha = \alpha^{\rm cr}$.  The two phases of the system
differ by the number of localized states. In the case
considered, this number is $N$ at $\alpha < \alpha^{\rm cr}$, 
and $N-1$ at  $\alpha > \alpha^{\rm cr}$. 

Much more interesting is the realistic case with the Hamiltonian 
(\ref{form5}). In this case,  trapping of resonance states occurs 
in the regime of overlapping resonances hierarchically, i.e.
one by one \cite{top}. The crossing points do {\it not} accumulate in
one point, but are distributed over a certain 
range of the parameter:  a {\it dynamical phase transition} takes
place in a finite parameter range inside the regime of overlapping 
resonances. It can therefore be observed \cite{top}. 
Also in this case, almost all resonance states are involved in the phase
transition of the system and, furthermore, the number $N$ of localized
states is reduced. 

The dynamical phase transition taking place in the system 
at high level density
causes finally a {\it dynamical stabilization} of the system: the system
consisting of only the localized long-lived  states 
beyond the phase transition  
is more stable than the  system below the dynamical phase
transition in spite of the stronger coupling between system and
environment beyond the phase transition.
The reason is the following: first the widths bifurcate  at
high level density and then the state with the shortest lifetime is
ejected. The dynamical stabilization is a global effect to
which all states contribute collectively by aligning one of the states 
(step by step) with a decay channel. In this manner, the sum of the
decay widths of the states of the system is reduced \cite{kleiro},
and the system is stabilized.

\section{Dynamical stabilization of different quantum systems 
in experimental data}
\label{7}

\subsection{Phase lapses}
\label{7a}

More than 10 years ago, in  experiments 
on Aharonov-Bohm rings containing a quantum dot in one arm, 
both the phase and the magnitude of the transmission amplitude
$T=|T|~e^{i\beta}$ of the dot are extracted \cite{laps1}. The results obtained
caused much discussion since they do not fit into the standard 
understanding of the transmission process.  
As a  function of the plunger gate voltage $V_g$, a series of well-separated 
transmission peaks of rather similar width and height has been observed
in many-electron dots
and, according to expectations, the transmission phases $\beta(V_g)$ 
increase continuously by $\pi$ across every resonance. 
In contrast to expectations, however, 
$\beta$ always jumps sharply downwards by $\pi$  in each valley
between any two successive peaks. These jumps called phase lapses, 
were observed 
in a large succession of valleys for every  many-electron dot studied.
Only in few-electron dots, the expected so-called mesoscopic behavior
is observed, i.e. the phases are sensitive to details of the dot 
configuration. The problem is considered theoretically in many papers 
over many years without solving it convincingly, e.g.  \cite{laps2}.

In \cite{laps3}, the generic features of phase lapses 
in the inelastic cross section are studied by using the toy model 
(\ref{sol15}) for the non-Hermitian Hamilton operator. 
According to the results of these calculations, 
the universal features observed in the phase lapses 
at high level density, in contrast to the mesoscopic features at low
level density, may be considered to be a hint at a dynamical phase transition. 
The transition occurs by controlling the system from low to high 
level density simulated in the calculations 
by means of $\alpha$. In accordance to this picture, only
the resonance states at low level density  show  
individual spectroscopic features. At high level density, the observed
resonances arise from trapped states. They show level repulsion, have 
vanishing spectroscopic relation to the open decay channels 
(i.e. small decay widths), and phase lapses appear. It follows further,
that any theoretical study on the basis of conventional Hermitian
quantum physics is unable to explain the experimental results
convincingly. More accurate calculations
on the basis of  (\ref{form6}) for the non-Hermitian Hamiltonian
$\heff$ are performed recently \cite{laps4} 
and compared with the experimental data. 

\subsection{Spin swapping operation}
\label{7b}

A swapping gate in a two-spin system exchanges the degenerate states 
$|\uparrow , \downarrow \rangle $ and $|\downarrow ,\uparrow \rangle $.  
Experimentally, this is achieved by turning on and off the spin-spin
interaction $b$ that splits the energy levels and induces an oscillation
with a natural frequency $\omega$. An interaction  $\hbar /\tau_{SE}$ 
with an environment of neighboring spins degrades this oscillation
within a decoherence time scale $\tau_\phi$. The experimental
frequency $\omega$ is expected to be roughly proportional to 
$b/\hbar$ and the decoherence time  $\tau_\phi$ proportional to
$\tau_{SE}$. In \cite{past}, experimental data are presented
that show drastic deviations in both $\omega$ and $\tau_\phi$ from
this expectation. Beyond a critical interaction with the environment,
the swapping freezes and the decoherence rate drops as $1/\tau_\phi
\propto (b/\hbar )^2 \tau_{SE}$. That means, the relaxation
decreases when the coupling to the environment increases.
The transition between these two
quantum dynamical phases occurs when $\omega  \propto
\sqrt{(b/\hbar)^2 -(k/\tau_{SE})^2}$ becomes imaginary (where $k$
depends only on the anisotropy of the system-environment interaction,
$0 \le k \le 1$). The experimental results are interpreted 
by the authors as an environmentally induced quantum dynamical
phase transition occurring in the spin swapping operation 
\cite{past}.

Further theoretical studies within the Keldysh formalism 
showed that $\tau_{\phi}$ is a non-trivial function of the 
system-environment interaction rate $\tau_{SE}$, indeed: it is 
$1/\tau_\phi \propto 1/\tau_{SE}$ at low  $\tau_{SE}$ 
(according to the Fermi golden rule) but
$1/\tau_\phi \propto \tau_{SE}$ at large $\tau_{SE}$. This theoretical 
result is in (qualitative) agreement with the experimental results.
In \cite{past}, the dynamical phase transition 
in the spin swapping operation is related to the
existence of an exceptional point. 

The dynamical phase transition observed experimentally in the spin 
swapping operation and described theoretically within the Keldysh formalism
shows qualitatively the same features as the dynamical phase transitions
discussed in the present paper on the basis of the resonance trapping
phenomenon (width bifurcation).

\subsection{Loss induced optical transparency in complex optical potentials}
\label{7c}

The optical wave equation for complex PT symmetric potentials
is formally equivalent to the quantum 
mechanical Schr\"odinger equation \cite{makris}.
One expects therefore that PT symmetric optical lattices 
show a behavior which is qualitatively similar to that
discussed  for open quantum systems. 

Experimental studies showed, indeed, a phase transition that leads to
a loss induced optical transparency in  specially designed
non-Hermitian guiding potentials \cite{opticsexp}: 
the output transmission first decreases, attains a minimum
and then increases with increasing loss. The phase
transition is related, in these papers, to PT symmetry breaking.
In a following theoretical paper \cite{makris2}, the Floquet-Bloch
modes are investigated in PT symmetric complex periodic
potentials. As a result, the modes are skewed (nonorthogonal) and
nonreciprocal. That means, they show the same features as  modes 
of an open quantum system under the influence of exceptional points.
A detailed discussion of this analogy is given in \cite{jopt}.

\subsection{Dicke superradiance and subradiance  in optics}
\label{7d}

The assumption that the probability of a given
molecule to emit a photon may be considered to be independent of the
states of the other molecules is justified only when 
the distance between the molecules is large.
Generally,  all the molecules are
interacting with the common radiation field  and the spontaneous
emission takes place coherently. Dicke \cite{dicke} was the first who 
considered the coherence in spontaneous radiation processes and, as a 
consequence, the formation of the so-called superradiant state.
The collective coupling of the atoms via the radiation field
leads also to a substantial radiative shift of the transition
energy, the so-called collective Lamb shift. This effect is recently
proven experimentally \cite{roehlsberger}. It will allow us to probe 
aspects of quantum
electrodynamics in relatively low-energy experiments \cite{scully}. 

By using the simple non-Hermitian Hamilton
operator (\ref{sol15}), the formation of a superradiant state can be
controlled by  the coupling strength $\alpha$ between system and
environment \cite{soze}.  At large $\alpha$, the short-lived
superradiant state  is formed together with long-lived
subradiant (trapped) states. The corresponding phase transition, called 
superradiance transition by the authors, is nothing but the dynamical phase
transition discussed in section \ref{6}. In both cases, the same Hamiltonian
with frozen internal degrees of freedom is used. In section \ref{6}
the relation to exceptional points is considered. 

Using an effective non-Hermitian Hamilton operator describing
photon-mediated atomic dipolar interactions, calculations
for the Dicke superradiance in atomic gases are
performed recently \cite{kaiser1}. The calculations show all the features
characteristic of a dynamical phase transition. Only the notations
used are different: for example 'disorder' used in \cite{kaiser1}
corresponds to 'individual spectroscopic properties of the states' in
the many-body problem (sections \ref{2} and \ref{3}).
Also in these calculations, the crossover from the uncorrelated to the
correlated behavior appears at a critical value of a certain
parameter. Interesting is the scaling behavior of the escape rates of
photons propagating in a 3D atomic gas when traced by means of a certain
parameter. For small parameter values, cooperative effects are
negligible, and photons are emitted in spontaneous and incoherent
processes. For larger values of this parameter, cooperative effects
set in and the Dicke superradiance appears.
The corresponding subradiance is considered in \cite{kaiser2}.  
It extends the lifetime of the excitation to many times the natural
lifetime of a single atom and is, therefore, doubtless interesting 
in quantum information science.

\section{Time in an open quantum system}
\label{8}

The results of sections \ref{4} to \ref{7}  show that
resonance states with very large widths, corresponding to very
short lifetimes, do not exist. Due to the avoided crossings of the
eigenvalue trajectories and the accompanying reduced phase rigidity 
$\rho < 1$, most resonance states are dynamically localized 
at high level density and have a finite non-vanishing lifetime. 
The following conclusions can be drawn.
\\
(i) Energy and time are determined by the eigenvalues of one and the same
operator, namely of $\heff$. The time operator is the non-Hermitian part 
of $\heff$ while the energy operator corresponds, as well known, 
to the Hermitian part of $\heff$.
\\
(ii) Not only time varies continuously (as well known) but also energy
does so, since most states of an open quantum system
are resonance states with a non-vanishing width.
\\
(iii) Time is bounded from below in a similar manner as energy.
\\
(iv) Discrete states correspond to $t \to \infty$. This relation
is analog to the assumptions of standard quantum mechanics described
by Hermitian operators.

The two operators Re$(\heff)$ and Im$(\heff)$ do not commute when the 
number $C$ of channels is different from the number $N$ of states, as
can be seen easily from (\ref{sol15}). Usually $C \ll N$. 
According to the points (i) to (iv), the
main argument by Pauli \cite{pauli} against the derivation of the time energy
uncertainty relation does not exist when an open quantum system
with the non-Hermitian Hamilton operator $\heff$ is considered.

The concept of time considered in the present paper
is characteristic of an open quantum system in the same manner as 
energy. It has a physical meaning only for $t_b < t < \infty$ 
where $t_b$ is the time at which width bifurcation creates the
dynamic stabilization of the system. The time $t$ defined in this manner
is a measurable quantity. Note $t$ is more sensitive to parameter
variations than $E$ (see \cite{roseb} and \cite{eleuch} for numerical 
examples).
 
The dynamical stabilization of the system at high level density
appears as a counterintuitive process when time is considered as a 
parameter that may be varied continuously between $-\infty$ and $+\infty$
and is  {\it not characteristic} of the system. In contrast to this, 
the finite value $t_b$ relies on the fact
that time is a value  {\it characteristic} of the system.
The dynamical stabilization occurs in consequence of the fact 
that the resonances will never really overlap.
The only way to achieve this is to accumulate almost all coupling
strength between system and environment onto one state 
(in the one channel case) while the
remaining states become stabilized (long-lived) \cite{dittes}.
The short-lived state created in this manner, is
aligned to the continuum of scattering states and 
does not belong to the set of localized states. 
Beyond $t_b$, the system differs from the
original one: the number of states is reduced and their individual
spectroscopic features are lost. Instead, cooperative effects 
are important (which may be disturbed only by coupling the system to
another channel). Mathematically, the redistribution
rests on the existence of exceptional points and the related
phenomenon of avoided level crossings. It occurs under the influence 
of the environment into which the system is embedded. 

Experimentally it is possible to form spatially remote
discrete states on separate
quantum point contacts and to allow an interaction between them via a common
continuum   \cite{bird}. The results show that the continuum supports an 
effective interaction between the two states which is 
mediated by the continuum and is highly robust: the detector exhibits
two Fano resonances due to the two different bound states.
These studies, revealing a pronounced avoided level crossing, 
show clearly that the continuum  affects the microscopic structure of
bound states in the two quantum point contacts.

In the examples discussed in section \ref{7}, the time $t_b$ appears 
in a natural manner. It causes  {\it measurable} effects. For example, the
transmission through a small quantum system is enhanced in the
parameter range in which the redistribution in the system takes place
\cite{burosa}. Interesting are the very stable whispering gallery modes
in a small quantum system. They are partly aligned 
to the scattering wavefunctions and their lifetimes are shorter
than those of the other states. They cause an enhancement of the
transmission, and the system becomes  almost transparent ($\rho
\to 0$)  under these conditions.  

The time $t_b$ determines also the brachistochrone problem \cite{top}
which consists of finding the minimal time for the transition from a
given initial state to a given final state of the considered system.  
At high level density, the individual resonance states can no longer 
be identified. Here $\rho < 1$, and  the wavefunctions of some
states of the system are partly aligned to the scattering
wavefunctions such that the time for the transition from a given
initial state to a given final state may be radically shortened. 
However, the time for traveling through the system does never
vanish. It is bounded from below since it can not be smaller than the
time corresponding to the transparency of the system  \cite{top}.  

The non-adiabatic processes found recently when cycling exceptional
points \cite{cycling}, are surely related to the finite time $t_b$ 
below which time loses its meaning in the system considered.
The cycling crosses regions with fundamental different time concepts.

\section{Summary}
\label{9}

In the present paper, the meaning of time in an open quantum system 
is considered. System as well as environment are quantum mechanical
objects with the consequence that the Hamilton operator $\heff$ 
of the system is non-Hermitian. The real parts of the eigenvalues of 
$\heff$ provide the energies of the states while the imaginary 
parts of them are related to their lifetimes. Time $t$ in the open
quantum system is defined by the lifetimes of the decaying states. 

In an open quantum system described by a non-Hermitian operator,
the main objection \cite{pauli} to the derivation of the time energy 
uncertainty relation does not occur. The time operator
appears in a natural manner together with the energy operator. Energy
{\it and} time are bounded from below and, furthermore, vary continuously
as a function of a parameter, as a rule. 

The open quantum system is reversible at low level density where the
levels are far from one another and the system can be described, to 
a good approximation, by a Hermitian operator. At high level density,
however, the levels avoid crossing and irreversible processes determine
the redistribution processes taking place in the system. These 
irreversible processes
are caused by nonlinearities in the Schr\"odinger equation of the open
system in the vicinity of avoided level crossings (and exceptional
points, respectively) due to the coupling of the states via the 
environment. A dynamical stabilization of the system occurs,
at the time $t_b$, under the influence of the environment. 
$t_b$ is the lowest value of $t$.

Further experimental as well as theoretical studies are necessary. 
Above all, the time energy uncertainty relation has to be derived
in a mathematical convincing manner for the general case with 
discrete and narrow resonance states of a many-particle system. 
Experimentally, the influence of a third state onto the mixing of
two states in the neighborhood of an avoided level crossing should
be studied. As discussed in appendix \ref{a5}, this is the basic process 
of the dynamical stabilization 
taking place in the system at high level density.

\section*{Appendix}

\vspace{.2cm}

\begin{appendix}

\section{The eigenvalues of a non-Hermitian $2\times 2$  operator}
\label{eiv}

Let us consider the  Hamiltonian 
\begin{eqnarray}
H(\omega) = \left(
\begin{array}{cc}
\epsilon_1 & \omega \\
\omega & \epsilon_2 
\end{array} \right) \, 
\label{int3}
 \end{eqnarray}
with the  energies $\epsilon_i ~(i=1,2)$ of the two states
and the interaction $\omega$ between them. The $\epsilon_i$ are
assumed to contain the corrections due to the coupling of the state $i$
to the environment. The eigenvalues are 
\begin{eqnarray}
\varepsilon_{1,2}&=&\frac{\epsilon_1 + \epsilon_2}{2} \pm Z \; ; \; \quad
Z=\frac{1}{2}\sqrt{(\epsilon_1 - \epsilon_2)^2 + 4 \omega^2} \; .
\label{int4}
\end{eqnarray}
The levels repel each other in energy according to the value
Re$(Z)$ while the widths bifurcate corresponding to Im$(Z)$.
The two eigenvalue trajectories cross when $Z=0$, i.e. when
$(\epsilon_1 - \epsilon_2)/2\omega = \pm \, i $.
At the crossing points, called mostly {\it exceptional points},
the two eigenvalues coalesce,
$\varepsilon_1 ~=~ \varepsilon_2 ~\equiv ~\varepsilon_0 $. 
In the vicinity of the crossing points, the dependence of the 
eigenvalue trajectories on the parameter is more complicated
than far from them: the two levels approach each other in energy
and the widths become equal so that
Re$(\varepsilon_1) \leftrightarrow {\rm Re}(\varepsilon_2)$ and
Im$(\varepsilon_1) \leftrightarrow {\rm Im}(\varepsilon_2)$
at the crossing point. 

When $H$ is a Hermitian operator, the unperturbed
energies $\epsilon_i $ of the states and the interaction $\omega$
between them are real. According to (\ref{int4}),
the two (real) eigenvalue trajectories 
$\varepsilon_i(\alpha)  = e_i(\alpha)$ 
cannot cross (for $\omega \ne 0$) when traced as a function of a 
certain parameter $\alpha$. Instead, they  avoid crossing. 
The fictive crossing point is called usually
{\it diabolic point}. The topological structure of this point is
characterized by the Berry phase \cite{berry}
which is studied theoretically and experimentally in many  papers.

The situation is another one when $H$ is a non-Hermitian operator. In such a
case, the unperturbed energies $\epsilon_i$ and also the interaction $\omega$
are complex, usually. The states  can decay, in general, and
the two eigenvalues  (\ref{int4}) can be written as
\begin{eqnarray} 
\varepsilon_{1,2}= 
e_{1,2} - \frac{i}{2} ~\gamma_{1,2}  \qquad ({\rm with} ~\gamma_{1,2}
\ge 0 ) \; .
\label{eiv1}
\end{eqnarray}
The widths $\gamma_i$ are proportional to the inverse lifetimes
$\tau_i^{-1}$ of the  states, $i=1,2$.
The two eigenvalue trajectories $\varepsilon_i(\alpha)$
may cross according to (\ref{int4}), and the
crossing point is an exceptional point in agreement with the 
definition given in \cite{kato}.
The topological phase of the exceptional point is  twice the Berry
phase \cite{top}. This theoretical result is proven experimentally
by means of a microwave cavity \cite{demb1}.
According to  (\ref{int4}), Re$(Z)$ causes repulsion of the levels in energy. 
It is the dominant part when the interaction $|\omega|$ of the states 
is small. The value Im$(Z)$  is  dominant
when $|\omega|$ is large. It is related, according to (\ref{int4}), 
to a bifurcation of the widths of the levels.  

The formal equivalence of the optical wave
equation in PT symmetric optical lattices to 
the quantum mechanical Schr\"odinger equation allows us to study the 
properties of quantum systems the states of which can not only decay 
due to their coupling to the environment according to (\ref{eiv1}),
but may also be formed out of the environment \cite{opticsexp}. 
In optics, these two possibilities are called {\it loss} and {\it gain}. 
In  PT symmetric optical lattices, the eigenvalues are \cite{jopt}
\begin{eqnarray}
\varepsilon_{1,2} = 
e_{1,2} \pm \frac{i}{2} ~\gamma_{1,2}  \qquad ({\rm with} 
~\gamma_{1,2} \ge 0 \; {\rm ~and} \; ~e_{1} = e_2) 
\label{eiv2}
\end{eqnarray}
in difference to (\ref{eiv1}). Due to PT symmetry, all 
eigenvalues $\varepsilon_i = e_i$ may be real 
(corresponding to $\gamma_i =0$) 
when Re$(Z) \gg$ Im$(Z)$, i.e. at low coupling of the states to the continuum. 
However, the PT symmetry is broken when  Im$(Z) \gg$ Re$(Z)$
and  $\gamma_{1,2} \ne 0$. 
The difference between the two models with the eigenvalues 
(\ref{eiv1}) and (\ref{eiv2}) will allow us to receive interesting 
information on quantum systems by
studying not only open quantum systems (which exist in nature) but
also  PT symmetric systems (which are formally
equivalent to them).

\section{The eigenfunctions of a non-Hermitian $2\times 2$ operator}
\label{eif}

The eigenfunctions of the non-Hermitian Hamilton operator $H$,
equation (\ref{int3}), are biorthogonal,
\begin{eqnarray}
\langle \phi_k^*|\phi_l\rangle  = \delta_{k, l} 
\; .
\label{eif1}
\end{eqnarray}
From these equations follows
\begin{eqnarray}
\langle \phi_k|\phi_{k }\rangle & \equiv &  A_k \ge 1 
\label{eif2}\\
\langle \phi_k|\phi_{l\ne k}\rangle =- 
\langle \phi_{l \ne k  }|\phi_k\rangle & \equiv & B_k^l ~; 
~~~|B_k^l|\ge 0 \; .
\label{eif3}
\end{eqnarray}
At the crossing point $A_k^{\rm (cr)} \to \infty,  
~|B_k^{l ~{\rm (cr)}}| \to \infty $, for details see \cite{top}. 

The relation between the eigenfunctions 
$\phi_1$ and $\phi_2$ of the operator (\ref{int3}) at the crossing 
point  is
\begin{eqnarray}
\phi_1^{\rm cr} \to ~\pm ~i~\phi_2^{\rm cr} \; ;
\quad \qquad \phi_2^{\rm cr} \to
~\mp ~i~\phi_1^{\rm cr}  
\label{eif5}
\end{eqnarray}  
according to analytical  as well as numerical studies \cite{top}.
That means, the state $\phi_1$ jumps, at the exceptional point, via
the chiral state ~$\phi_1\pm i\phi_2$ ~to the state  ~$\pm i \phi_2$.

The two eigenfunctions are linearly dependent of one another at the
crossing point such that the number of eigenfunctions of $H$
seems to be reduced at this point. 
Theoretical studies \cite{gurosa} have shown however that 
{\it associated vectors} $\phi_i^{cra}$ defined by the Jordan relations, 
appear at the crossing points. The corresponding equations are
\begin{eqnarray}
(H -\varepsilon_0)~\phi_{1,2}^{\rm cr} = 0 \; ; \quad \quad 
(H -\varepsilon_0)~\phi_{1,2}^{\rm cra} = \phi_{1,2}^{\rm cr} \; .  
\label{eif8}
\end{eqnarray}
The existence of two states in the very neighborhood of the 
exceptional point has
been seen in a numerical calculation for the elastic scattering of a
proton on a light nucleus \cite{plo}: the elastic scattering phase
shifts  jump always by
$2\pi$ (and not by $\pi$ as for a single resonance state).

In an experimental study on a microwave cavity \cite{demb1}, the topological
structure of the exceptional point and its surrounding is studied by
encircling it and tracing the relative amplitudes of the 
wavefunctions (field distributions inside the cavity). As a result,  
the wavefunctions including their phases are restored after four
surroundings. The authors \cite{demb1} interpreted the experimental
data by two theoretical assumptions: (i)
the two wavefunctions coalesce into one at the exceptional point,
$\phi_1^{\rm cr} \leftrightarrow \phi_2^{\rm cr} $, and (ii)
only one of the wavefunctions picks up a phase of $\pi$ (a sign change) 
when encircling the  critical point. 
The experimental result  can be explained, however, without any
additional assumptions by using the relations (\ref{eif5}):\\
\hspace*{.5cm} 1. cycle: \hspace*{.2cm}$\varepsilon_{1,2} \to \varepsilon_{2,1}; 
 \hspace*{.45cm} \phi_{1,2} \to \pm\,i\,\phi_{2,1}$~; \hspace*{.5cm}
2. cycle: \hspace*{.2cm}$\varepsilon_{2,1} \to \varepsilon_{1,2};
\hspace*{.3cm}\pm\, i\,\phi_{2,1} \to -\phi_{1,2}$\\
\hspace*{.5cm} 3. cycle: \hspace*{.2cm}$  \varepsilon_{1,2} \to 
\varepsilon_{2,1};     
\hspace*{.2cm}- \phi_{1,2} \to \mp \, i\,\phi_{2,1}$~; \hspace*{.5cm}
4. cycle: \hspace*{.2cm}$ \varepsilon_{2,1} \to \varepsilon_{1,2};     
\hspace*{.3cm}\mp \, i  \,\phi_{2,1} \to ~\phi_{1,2}$ .\\  
The eigenvalues are restored after two surroundings
and  the eigenfunctions are restored  after four
surroundings,  in full agreement with the experimental result.
In any case, $|~\phi_1^{\rm cr}~|~ = ~|~\phi_2^{\rm cr}~| $
at the crossing point.
The topological phase is twice the Berry phase, in accordance with the
enlarged function space in open quantum systems.

Furthermore, the phases of the wavefunctions jump by $\pi /4$ 
at the crossing point (when traced as a function of a parameter) 
due to the biorthogonality  (\ref{eif1})
of the eigenfunctions of the non-Hermitian Hamiltonian $H$,
see also (\ref{eif2}) and (\ref{eif3}).
This result has been proven in many numerical studies, see \cite{top}.

\section{The phase rigidity $r_\lambda$ of the eigenfunctions of a
  non-Hermitian $2\times 2$  operator}
\label{a3}

Let us now consider the consequences of the biorthogonality relations
(\ref{eif1}) and (\ref{eif2})
for the two borderline cases characteristic of neighboring resonance states.\\  
(i) The two levels are distant from one another. Then the eigenfunctions
 are (almost) orthogonal,  
$\langle \phi_k^* | \phi_k \rangle   \approx
\langle \phi_k | \phi_k \rangle  = A_k \approx 1 $.\\
(ii) The two levels cross. Then the two eigenfunctions are linearly
dependent according to (\ref{eif5}) and 
$\langle \phi_k | \phi_k \rangle = A_k \to \infty $.\\
These two relations 
show that the phases of the two eigenfunctions
relative to one another change when the crossing point is approached. 
This can be expressed quantitatively by defining the {\it phase
  rigidity} $r_k$ of the eigenfunctions $\phi_k$,
\begin{eqnarray}
r_k ~\equiv ~\frac{\langle \phi_k^* | \phi_k \rangle}{\langle \phi_k 
| \phi_k \rangle} ~= ~A_k^{-1} \; . 
\label{eif11}
\end{eqnarray}
It holds $1 ~\ge ~r_k ~\ge ~0 $.  
The  non-rigidity $r_k$ of the phases of the eigenfunctions of $H$ 
follows also from the fact that $\langle\phi_k^*|\phi_k\rangle$
is a complex number (in difference to the norm
$\langle\phi_k|\phi_k\rangle$ which is a real number) 
such that the normalization condition
(\ref{eif1}) can be fulfilled only by the additional postulation 
Im$\langle\phi_k^*|\phi_k\rangle =0$ (what corresponds to a rotation). 

Note, discrete states never cross but always avoid crossing, see
appendix \ref{eiv}. The eigenfunctions are real and normalized as 
$\langle \phi_k|\phi_k\rangle = 1$ (see appendix \ref{eiv}). 
Accordingly, $r_k=1$ for discrete states also in the region of an avoided 
level crossing. The corresponding crossing (exceptional) point can be
found by analytical continuation into the continuum \cite{solov}.
 
The variation of $r_k$ in approaching the crossing point of two 
eigenvalue trajectories 
of resonance states is proven experimentally by means of a
study on a microwave cavity \cite{demb2}. As a result, 
the phase difference between two modes is $\pi$ at large distance and
decreases to $\pi /2$ at the crossing point.
The authors of \cite{demb2} interpret the experimental data
by assuming (i) that the singular point is  a chiral state 
(in spite of the phase jump occurring at the crossing point,
when traced as a function of a certain parameter, see 
(\ref{eif5})),
(ii) that the number of states is reduced from 2 to 1 at the crossing
point (in spite of the existence of the associate vector (\ref{eif8}))
and (iii) that a single point in the continuum can be identified
(although it is of measure zero).
The authors are unable to explain  the large parameter range 
in which the phase difference decreases in approaching the crossing point.

Considering the phase rigidity $r_k$ in the regime of the two
overlapping resonance states, no  additional assumptions are required 
for the explanation of the experimental results given in \cite{demb2},
since the phase rigidity (being a quantitative measure for the degree 
of resonance overlapping) varies smoothly in a comparably large 
parameter range. It can therefore be concluded that the experimental
results  \cite{demb2}
prove the statement that the phases of the eigenfunctions of the
non-Hermitian Hamilton operator $H$, equation (\ref{int3}),  
are not rigid in approaching the crossing point.

\section{Nonlinear source term in the Schr\"odinger equation in the
  neighborhood of an exceptional point}
\label{a4}

According to (\ref{int3}), the Schr\"odinger equation with the 
unperturbed operator 
$H_0\equiv H(\omega = 0)$ and a source term
arising from the interaction $\omega$ with another state reads \cite{ro01}
\begin{eqnarray}
\label{mix1}
(H_0  - \epsilon_n) ~| \phi_n \rangle & = &
- \left(
\begin{array}{cc}
0 & \omega \\
\omega & 0 
\end{array} \right) |\phi_n \rangle 
\equiv  W ~| \phi_n \rangle \nonumber \\
&=& \sum_{k=1,2} \langle \phi_k |W|\phi_n\rangle
\{ A_k ~|\phi_k\rangle + 
\sum_{l\ne k} ~B_k^l ~|\phi_l\rangle \} \; .
\end{eqnarray}
Here
$\langle \phi_k|\phi_{k }\rangle  \equiv   A_k  \ge 1 $ according to 
(\ref{eif2}) and $
\langle \phi_k|\phi_{l\ne k }\rangle = - 
\langle \phi_{l \ne k  }|\phi_{k}\rangle  \equiv 
B_k^{l}, ~|B_k^{l}|\ge 0 
$ according to  (\ref{eif3}). The $A_k$ and $B_k^l$
characterize the degree of resonance overlapping.
In the regime of overlapping resonances, $1>A_k >0$, $|B_k^l| >0$, and 
equation (\ref{mix1}) is nonlinear. 
The most important part of the nonlinear contributions is contained in 
\begin{eqnarray}
\label{mix2}
(H_0  - \epsilon_n) ~| \phi_n \rangle =
\langle \phi_n|W|\phi_n\rangle ~|\phi_n|^2 ~|\phi_n\rangle  
\end{eqnarray}
which is a nonlinear Schr\"odinger equation. According to
(\ref{mix1}), the nonlinear Schr\"odinger equation (\ref{mix2})
passes smoothly into the standard linear Schr\"odinger equation
when $A_k \to 1$ and $B_k^l \to 0$.

\section{Time reversal symmetry breaking in the neighborhood of an
  exceptional point}
\label{a5}

Exceptional points that are well separated from the influence of
external sources (including the influence caused by
other resonance states), are highly symmetric in
approaching them. That means, the two states pass one into the other one
according to (\ref{eif5}) with an exchange of their wavefunctions,
$\phi_1^{\rm cr} \to ~\pm ~i~\phi_2^{\rm cr}$ and $ \phi_2^{\rm cr} \to
~\mp ~i~\phi_1^{\rm cr}$. At a certain finite distance from the
exceptional point, there are again two states with the wavefunctions 
$|\phi_1|$ and  $|\phi_2|$, respectively.

This symmetry may be distorted under the influence of an external
magnetic field as has been shown experimentally on a microwave 
cavity \cite{trsb}.
The magnetic field causes time reversal symmetry breaking. 

The symmetry may be disturbed also by the influence of
another resonance state in the
neighborhood due to the finite parameter range around the exceptional
point  in which the 
wavefunctions of the two states are mixed  with each other \cite{ro01}. 
When the interaction of the third state is symmetric relative to
the two crossing ones, the third state will appear 
as an {\it observer} and time reversal symmetry is not broken.
Numerical examples of such a situation are shown in the transmission 
through a quantum dot \cite{trsbsadreev} and also in the generic case
studied in \cite{eleuch}.
When the interaction of the third state  with the two
crossing ones is, however, not symmetrically, 
time reversal symmetry may be broken and may cause irreversible
processes due to the nonlinear terms in the Schr\"odinger equation as
discussed in appendix \ref{a4}.

It would be highly interesting to study 
experimentally time reversal symmetry breaking in
the case of an exceptional point disturbed non-symmetrically by a third
state in the neighborhood. 

\end{appendix}

\end{document}